\documentclass[prb,twocolumn,
superscriptaddress,longbibliography]{revtex4-2}
\usepackage{pdfpages}

\usepackage{amsthm}
\usepackage{amssymb}
\usepackage{booktabs}
\usepackage[english]{babel}
\usepackage{color}
\usepackage{comment}
\usepackage{hyperref}
\usepackage[utf8]{inputenc}
\usepackage[fleqn]{mathtools}
\usepackage{sidecap}
\usepackage{svg}
\usepackage{capt-of}
\usepackage{xr}

\makeatletter
\AtBeginDocument{\let\LS@rot\@undefined}
\makeatother

\makeatletter\AtBeginDocument{\let\@elt\relax}\makeatother

\newcommand{\rev}[1]{#1}

\newcommand{\manuscripttitle}{Physics-inspired Equivariant Descriptors of Non-bonded Interactions}

\theoremstyle{definition}

\theoremstyle{remark}

\newcommand{\br}{\boldsymbol{r}}

\newcommand{\hr}{\hat{\br}}

\newcommand{\intcutoff}{\int_0^{r_\mathrm{cut}}\mathrm{d}^3\br\,}

\makeatletter
\newcommand*{\addFileDependency}[1]{%
\typeout{(#1)}
\@addtofilelist{#1}
\IfFileExists{#1}{}{\typeout{No file #1.}}
}\makeatother

\newcommand*{\myexternaldocument}[2][]{%
\externaldocument[#1]{#2}%
\addFileDependency{#2.tex}%
\addFileDependency{#2.aux}%
}

\myexternaldocument[si-]{si}
\begin{document}

\title{\manuscripttitle}
\author{Kevin K. Huguenin-Dumittan}
\altaffiliation{Contributed equally to this work}
\author{Philip Loche}
\altaffiliation{Contributed equally to this work}
\author{Ni Haoran}
\author{Michele Ceriotti}
\email{michele.ceriotti@epfl.ch}
\affiliation{Laboratory of Computational Science and Modeling, IMX, École Polytechnique Fédérale de Lausanne, 1015 Lausanne, Switzerland}

\begin{abstract}
\rev{One essential ingredient in many machine learning (ML) based methods for atomistic modeling of materials and molecules is the use of locality. 
While allowing better system-size scaling, this systematically neglects long-range (LR) effects, such as electrostatics or dispersion interaction. We present an extension of the long distance equivariant (LODE) framework that can handle diverse LR interactions in a consistent way, and seamlessly integrates with preexisting methods by building new sets of atom centered features.
We provide a direct physical interpretation of these using the multipole expansion, which allows for simpler and more efficient implementations. 
The framework is applied to simple toy systems as proof of concept, and a heterogeneous set of molecular dimers to push the method to its limits.
By generalizing LODE to arbitrary asymptotic behaviors, we provide a coherent approach to treat arbitrary two- and many-body non-bonded interactions in the data-driven modeling of matter.}
\end{abstract}
\maketitle

Modeling approaches based on machine learning (ML) have become ubiquitous in the field
of atomistic simulations, bridging the gap between techniques purely based on classical
mechanics and empirical forcefields, which are fast but less accurate, and quantum
mechanical approaches providing greater accuracy at a larger cost \cite{
behler_generalized_2007, bartok_gaussian_2010, bartok_representing_2013,
musil_physics-inspired_2021}. Since the introduction of early ML models to predict the
energy of extended atomic structures \cite{behler_generalized_2007,
bartok_gaussian_2010} and molecules\cite{rupp+12prl}, the field has rapidly expanded in
many directions, improving the accuracy of predictions, and extending the diversity of
target properties beyond energies and forces, including vectorial and tensorial
properties \cite{glie+17prb,grisafi_symmetry-adapted_2018,wilkins_accurate_2019},
electron densities \cite{broc+17nc,alre+18cst,fabrizio_electron_2019,
lewis_learning_2021}, single-particle electronic Hamiltonians
\cite{schu+19nc,nigam_equivariant_2022} and many-particles
wavefunctions\cite{carl-troy17science,herm+20nc}. A key ingredient in most successful
approaches has been the use of locality, often justified in terms of the
``nearsightedness'' of electronic structure \cite{kohn_density_1996,
prodan_nearsightedness_2005}. Local models truncate atomic interactions up to a cutoff
radius, which allows the development of fast algorithms scaling linearly with the number
of particles.

Introducing a cutoff, however, neglects important contributions from all sorts of
long-range (LR) interactions. The most prominent LR effects are the $1/r$ Coulomb
potential between electrical charges as well as the $1/r^6$ dispersion interactions
between induced dipoles \cite{ambr+16science}. Many other asymptotic decays exist, e.g.
charge-dipole and hydrogen bonding ($1/r^2$), permanent dipole-dipole ($1/r^3$), and
charge-non-polar ($1/r^4$)\cite{jackson_classical_1998} interactions. On a
coarser-grained scale, the effective potential in dense polymer solutions, between
particles and surfaces, or membranes in biological systems involves different
combinations of exponents \cite{israelachvili_intermolecular_2011,
kanduc_attraction_2014}.

At the simplest level, neglecting LR effects sets a lower bound to the possible
prediction error. In some cases, LR interactions determine qualitatively different
behavior in materials \cite{hansen_theory_2013}. In order to address these issues in
machine learning models, several approaches have been proposed. If the goal is to
predict the energy and the forces of an atomic structure, the simplest strategy is to
add an explicitly-fitted $1/r^p$ potential baseline. This idea was already applied to
one of the earliest ML potentials \cite{bartok_gaussian_2010}, and can easily be
combined with any ML scheme \cite{deng_electrostatic_2019,deringer_general-purpose_2020,
niblett_learning_2021}. Such a simple approach, in which each chemical species is
assigned a fixed point charge, has clear limitations, and more sophisticated techniques
have been proposed. One direction is to explicitly include the Wannier centers
associated with the valence electrons into the ML framework \cite{peng_efficient_2023,
zhang_deep_2022}. Given that the center of charge of the electrons can be different from
the positions of the nuclei, these models can describe both ionic charge and local
polarization. In order to describe global charge transfer, without violating
charge-neutrality constraints, several methods have been proposed that employ a global
charge equilibration scheme, predicting electronegativities rather than the charges
\cite{faraji_high_2017, ko_fourth-generation_2021, ko_general-purpose_2021}, and more
generally self-consistent treatment of electrostatics\cite{gao-rems22nc}. While the
Coulomb potential has been at the center of these developments, some of these methods
have also been extended to other interactions, including dispersion
\cite{deringer_general-purpose_2020} and more general potentials
\cite{anstine_machine_2023}, often using a rather explicit physics-based functional
form. There is, however, a lack of a unifying framework that treats various types of LR
interactions consistently.
One promising approach to \rev{fill in this gap} is the Long-Distance Equivariant (LODE)
framework\cite{grisafi_incorporating_2019, grisafi_multi-scale_2021}, that encodes LR
structural data in a form that mimics the asymptotic behavior of electrostatic
interactions. Thus, even if LODE features are ``physics-inspired'' -- and can be related
to explicit physical interactions when used in a linear model --  they retain the full
flexibility of general ML schemes.

Here, we generalize LODE features to mimic the
asymptotic behavior of any potential with an inverse power law form beyond Coulomb
interactions, and provide an efficient implementation that also includes gradients of
the descriptors with respect to atomic positions, that are needed to compute forces. We
provide a detailed mathematical analysis that allows an exact physical interpretation of
the resulting LODE coefficients, \rev{and explain how the framework can also describe
many-body effects beyond pair potentials.} We finally investigate the subtle balance
between physical interpretability and generality of the ML model using two families of
datasets: simple toy systems and a diverse collection of molecular dimers.

\begin{figure}
    \centering
    \includegraphics[width=0.5\textwidth]{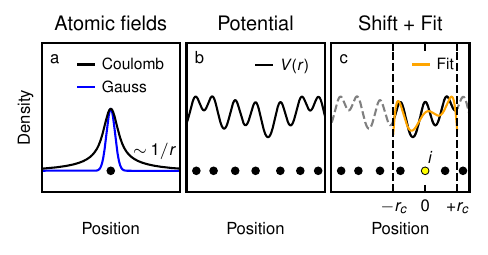}
    \caption{\rev{Construction of the LODE features. (a) shows a Gaussian density and the
    Coulomb potential generated by it, which is a smeared out version of the bare $1/r$
    Coulomb potential. The key for the LODE construction is the significantly slower
    decay of the smeared Coulomb potential. (b) shows the potential field $V(r)$
    obtained as the superposition of the smeared Coulomb potential on all atoms (black
    circles). In panel (c), $V_i$ is defined as the restriction of $V$ to a local
    environment around the position of atom $i$ (yellow circle), where the coordinate
    system has also been shifted. $V_i$ is then discretized using atom-centered basis functions. The
    resulting coefficients, obtained from \autoref{V_projection}, are the LODE
    features.} }
    \label{fig:lode-overview}
\end{figure}

\rev{We first provide a concise summary on the construction of the LODE descriptors as
is discussed in Refs.~\citenum{gris-ceri19jcp} and \citenum{gris+21cs}, and illustrated
in \autoref{fig:lode-overview}. A self-contained and more detailed discussion can be
found in \autoref{si-sec:theory} in the Supporting Information.} The position of the
atoms in a structure is encoded in a permutation-invariant way by defining a smooth atom
density $\rho(\br)$ -- the same that is used to define local atom-density descriptors
including the popular smooth overlap of atomic positions (SOAP) descriptor
(\autoref{fig:lode-overview}a). The Coulomb potential $V(\br)$ generated by this density
(\autoref{fig:lode-overview}b) can then be computed efficiently (e.g. with an Ewald
summation in reciprocal space\cite{nijboer_calculation_1957}). \rev{The atom centered
features for an atom $i$ are then generated by first shifting the coordinate system to
its position $\br_i$, leading to the atom-centered potential $V_i(\br) = V(\br_i+\br)$.
This potential, evaluated up to some cutoff radius $r_\mathrm{cut}$, is then projected
onto a set of basis functions consisting of (real) spherical harmonics $Y_l^m$ specified
by the angular indices $l=0,1,\dots,l_\mathrm{max}$ and $|m|\leq l$, as well as radial
basis functions $R_{nl}(r)$ for $n=0,1,\dots,n_\mathrm{max}-1$ via the integral
    \begin{align}\label{V_projection}
        V_{i,nlm} & = \intcutoff R_{nl}(r)Y_l^{m}(\hr)V_i(\br),
    \end{align}
where $\br$ corresponds to the displacement from atom $i$.} Due to the slow decay of the
$1/r$ potential, these LODE coefficients contain information on the position of far-away
atoms, despite using an environment cutoff for the integration. In fact, similarly to
a Fourier expansion, knowing all coefficients in the limit as $n_\mathrm{max},
l_\mathrm{max}\rightarrow \infty$ would allow one to recover the original function
$V_i$, meaning that the LODE coefficients also play a role of fitting coefficients as
shown in \autoref{fig:lode-overview}c. They have the same mathematical properties as
those used to discretize the local density $\rho$, and can be used in a similar way,
combining them in a symmetry-adapted fashion to obtain invariant features analogous to
the SOAP \cite{bartok_gaussian_2010, bartok_representing_2013} descriptor and its
higher-order invariant and equivariant extensions \cite{grisafi_symmetry-adapted_2018,will+19jcp,
nigam_equivariant_2022, drautz_atomic_2019,musil_physics-inspired_2021}. In particular,
it was shown that a combination of local density coefficients and Coulomb-field LODE
leads to ``multi-scale'' features, that, when used in linear models, can be interpreted
in relation to the multipole expansion of the electrostatic potential
\cite{grisafi_multi-scale_2021}.

\rev{Aa a first key result in this work, for which we provide a detailed derivation in sections \ref{si-sec:multipole} and \ref{si-sec:theory} in
the Supporting Information, we present an exact physical interpretation of the LODE
cofficients $V_{i,nlm}$. For a fixed center atom $i$, the potential $V_i^>(\br)$ around
atom $i$ generated by the charge density \emph{outside} the cutoff region can be shown
to be of the form
\begin{equation}\label{multipole_main}
    V_i^>(\br) = \sum_{lm} M^>_{i,lm} r^l Y_l^m(\hat{\br}),
\end{equation}
where the coefficients $M^>_{i,lm}$ are called (exterior) multipole moments and
completely characterize the potential generated by the exterior atoms as discussed in more detail in \autoref{si-sec:multipole} of the Supporting Information. The main result
here is that with an appropriate choice of radial basis, the LODE coefficients are
precisely \emph{equal} to the multipole moments, $V_{i,nlm} = M^>_{i,lm}$, and in
particular contain the full information on the exterior atoms for what concerns
electrostatic properties.} Furthermore, our derivation shows that one does not need a
large number $n_\text{max}$ of basis functions, since \autoref{multipole_main} does not
depend on an index $n$. By choosing $R_{0l}(r)=r^l$, a single monomial basis function
per angular momentum channel $l$ is sufficient to capture the electrostatic
contributions from far-field atoms. This translates in computational savings by a factor
of $n_\text{max}$, which varies between 4 and 12 in typical applications. \rev{We will
call this the monomial or optimized basis, since this is the most compressed form in
which we can store the exterior information.}

A potential risk in previous LODE implementations
\cite{grisafi_incorporating_2019,grisafi_multi-scale_2021}, that are fine-tuned to
capture electrostatic behavior, is that they will be less flexible in describing other
long-range interactions. \rev{To address this issue, we generalize the LODE construction
to arbitrary $1/r^p$ interactions, leading to a $p$-dependent family of potentials
$V^{(p)}$. While conceptually straightforwad, this change does come with several
subtleties.} First, a key ingredient to obtain well defined coefficients in the
Coulombic case was the use of Gaussian charge densities rather than point charges to
remove the singularity of $1/r$ at the origin. For $p \geq 3$, the divergence as
$r\rightarrow 0$ becomes so strong that even using a Gaussian smearing, the resulting
potential is still singular. \rev{Thus, a family of effective potentials parametrized by
$p$ and having the correct $1/r^p$ behavior for large distances is used instead of the
bare form\cite{nijboer_calculation_1957, williams_accelerated_1971,
williams_accelerated_1989}.} The second difference is that the multipole expansion for a
$1/r^p$ potential with $p\neq 1$ contains several terms that do not appear in the
Coulombic case. Despite these additional complications, all results can still be translated
to the general case with suitable modifications. Both of these subtleties are discussed in
\autoref{si-sec:density_contribution} in the Supporting Information.
\rev{While these results apply to interactions that can be treated as pair potentials, the LODE framework can also describe a wide range of LR many-body interactions beyond pair potentials by combining multiple LODE coefficients to generate higher order invariants. 
This approach, which we discuss in more detail in \autoref{si-sec:manybody}, leads to the LR analogue of systematic body-order expansions used in methods based on atom-centered density correlations\cite{will+19jcp}, including ACE \cite{drautz_atomic_2019} or NICE \cite{niga+20jcp}.}

To assess the capabilities of the general LODE framework, we begin with a demonstrative
example, using a modified version of the toy dataset originally presented in Ref.
\citenum{grisafi_multi-scale_2021}. It consists of 2000 structures, each obtained by
distributing at random 64 particles inside cubic cells of varying size, such that no
particles are closer than 2.5\,Å. For the same particle positions we consider two
different potentials: in one case, similar to Ref.~\citenum{grisafi_multi-scale_2021},
we treat the structures as an overall neutral cloud of $\pm 1$ charges. In the second
test, all atoms are equal, and interact via an attractive $1/r^6$ dispersion
interaction. Periodic boundary conditions are used in both cases. 

\begin{figure}
    \centering
    \includegraphics[width=0.45\textwidth]{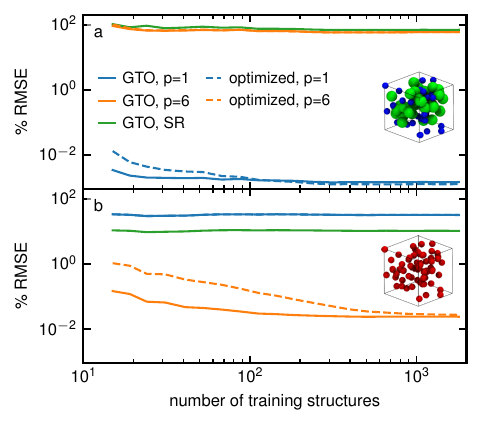}
    \caption{ \rev{ Learning curves showing the validation \%\,RMSE against the number
    of training structures. In (a), half of the the particles each have a charge of $\pm
    1$, while in (b), all particles are neutral and interact solely via an attractive
    dispersion interaction. Solid lines show models using the GTO basis with
    $n_\text{max}=8$ coefficients, whereas dashed lines use the optimized basis with a
    single coefficient. To convert back to absolute errors, the standard deviations in
    the energies are $\sigma=1.15$\,eV/atom and $\sigma=0.14$\,eV/atom, respectively.}}
    \label{fig:pointcharges}
\end{figure}

\rev{The energies of these systems are learned using linear models built either on
short-range (SR) or LODE descriptors with $p=1$ and $p=6$. The latter are computed with
a method similar to Ewald summation using a reciprocal space sum, matching the
periodicity of the target system. For the LODE descriptors, we only take the
coefficients for which $l=m=0$, and compare two choices of radial basis functions
$R_{nl}$. As a baseline, we use the Gaussian-type orbital (GTO) basis functions used in
Ref.~\citenum{grisafi_multi-scale_2021} as well as for various SR models including this
test, with $n_\mathrm{max}=8$ radial basis functions. This is compared with the optimal
radial basis that uses a single basis function. In fact, for $l=m=0$, both the spherical
harmonic $Y_0^0 = 1/\sqrt{4\pi}$ and $R_{0l}=r^l=1$ are just constant functions. Thus,
it can be seen from \autoref{V_projection} that the  coefficient $V_{i,000}$ that enters
the model is, up to a global factor,  the average within the cutoff sphere of the
potential generated by the atom density. For sufficiently small cutoff radii, we
therefore simply recover the potential at the position of atom $i$. Model details and
parameters can be found in \autoref{si-sec:model_pointcharges} of the Supporting
Information.}

The results in \autoref{fig:pointcharges} show the percentage room mean squared error
($\%$RMSE), defined as the absolute RMSE divided by the standard deviation $\sigma$ of
the energy in the training set, against the number of training set structures. We
observe that the SR descriptors, even when using a large $9$\,Å cutoff, are unable to
learn electrostatic or dispersion interactions, leading to rapidly-saturating learning
curves. On the other hand, the results show clearly how using a generalized LODE
descriptor adapted to the asymptotic decay of the LR potential allows to learn the
target potential with very high accuracy, and very few training structures. By using the
optimized radial basis, a single LODE coefficient can reach the same accuracy of a model
using an expansion on $n_\text{max}=8$ GTO basis functions.

In light of the analytical results discussed above, these observations are unsurprising.
In fact, it is more insightful to comment on the residual errors. First, it is clear
that the choice of a physically constrained model affects its applicability to other
types of interactions. The performance of a $V^{(1)}$ linear model for a $1/r^6$
potential, or those of a $V^{(6)}$ for $1/r$ are comparable or worse to those of a SR
model. Second, interactions-adapted models achieve high accuracy, but there is a
residual error, which is due to the finite Gaussian smearing of the atoms and
convergence of the reciprocal space sum in the implementation. While Ewald-based methods
in classical molecular dynamics, including PME \cite{darden_particle_1993},
SPME\cite{essmann_smooth_1995} and
P3M\cite{eastwood_particle-particleparticle-mesh_1988} correct for the smearing by using
a compensating SR part \cite{frenkel_understanding_2002}, the LODE descriptor itself
does not contain such corrections. This would not be an issue in practice, however,
since one typically combines LODE together with SR representations for more complex
systems. The slightly slower convergence of the optimized radial basis can also be
explained by the finite Gaussian smearing, and the cutoff radius for the integration in
Equation \eqref{V_projection}, leading to discrepancies between the true potential and
the LODE coefficients that can be compensated with a sufficient amount of data and/or
fitting coefficients.

\begin{figure*}
    \centering
    \includegraphics[width=1.0\textwidth]{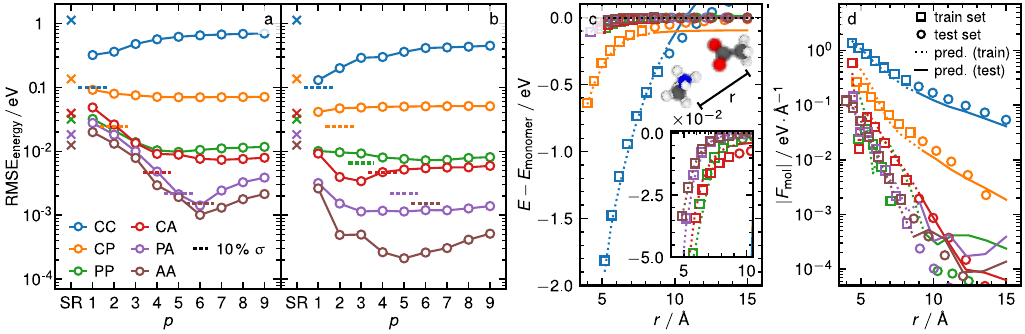}
    \caption{ \rev{ (a) Energy RMSE for for models using a single monomial basis as a
            function of the potential exponent $p$ for the six different dimer classes.
            Lines between the open circles serve as a guide to the eye. The dashed
            horizontal lines indicate 10\% of the standard deviation $\sigma$ of the
            training energies for each subset, which indicates the energy scale of each
            type of interactions. All lines are color-coded based on the type of dimers,
            as indicated in the legend. (b) Same as in panel (a) but using multiple
            radial basis functions ($n_\text{max}=6, l_\text{max}=4$). (c)~Binding
            energies $E-E_\mathrm{monomer}$ as a function of the distance $r$. Open
            squares show data taken for the train set while open circles show data from
            the test set. Lines show the predictions of the model. For each dimer class,
            the exponent $p$ with the highest prediction accuracy in panel (b) is used.
            The representative dimer is chosen such that its prediction errors are
            closest to the average behavior. The inset shows a snapshot of a positively
            charged methylamine and negatively charged acetate at a center of mass
            distance of $r_0=4.6$\,Å. (d)~Molecular force $\vert F_\mathrm{mol} \vert$
            of the first molecule as a function of the distance $r$. Symbols and curves
            are chosen in the same manner as in panel (c).}}
    \label{fig:rmse_subset}
\end{figure*}

From this first example, we conclude that a single $l=0$ component of the extended LODE
framework (1) is sufficiently expressive to learn a LR potential with matching exponent
with high accuracy and (2) has an increased efficiency due to a better choice of radial
basis. \rev{Beyond such pair or two-body potentials, we also show in
\autoref{si-sec:manybody} how higher order invariants built from the LODE coefficients
can be used to learn three-body dispersion, supporting the more general result that the
framework is also capable of treating many-body LR effects.} Armed with descriptors that
have a rigorous physical interpretation, but can be easily combined with arbitrary ML
frameworks, we can now explore the interplay between descriptors, model architecture,
and type of LR interactions \rev{using a more challenging dataset, with target
properties obtained from actual quantum mechanical methods.}

To cover a wide range of interactions between sufficiently simple but relevant
    molecules, we base our tests, on the BioFragment Database (BFDb)
    \cite{burns_biofragment_2017, grisafi_multi-scale_2021}, that contains 2291 pairs
    created from 22 relaxed organic molecules. For each configuration, we evaluate
    energies and forces along binding curves built starting from the configuration
    included in the BFDb with initial separation $r_0$, and increasing the separation
    along the line connecting the centers of mass (COM) of the two molecules, up to
    $r=15$\,Å between the COMs. We note that (1) we use the HSE06 hybrid functional
    \cite{heyd_hybrid_2003} with a high fraction of exact exchange, to reduce the DFT
    localization error and ensure that charged dimers dissociate without spurious
    fractional charges, (2) we include a non-local many-body dispersion
    correction\cite{hermann_density_2020} and (3) we use a supercell approach, so that
    binding energies also contain interactions between periodic replicas, consistent
    with the reciprocal-space calculation of LODE descriptors. Full details for the
    dataset construction and the models are provided in \autoref{si-sec:dimer_system} of
    the Supporting Information.

The BFDb contains charged molecules, polar molecules with \rev{an effectively constant}
dipole moment, and apolar molecules without any permanent charge or dipole moment. The
combination of these three molecular categories results in six dimer classes with
different ideal power-law decay constants of their interactions
\cite{israelachvili_intermolecular_2011}: charge-charge (CC) with $p_\mathrm{ CC}= 1$,
charge-polar (CP) with $p_\mathrm{ CP} = 2$, polar-polar (PP) with $p_\mathrm{ PP} = 3$,
charge-apolar (CA) with $p_\mathrm{ CA} = 4$, polar-apolar (PA) with $p_\mathrm{ PA} =
5$, and finally apolar-apolar (AA) with an ideal interaction decay of $p_\mathrm{ AA} =
6$. We show an example snapshot for a CC pair in the inset of \autoref{fig:rmse_subset}c
and example energy and force binding curves are shown in
\autoref{si-fig:dft_dimer_curves_fit} of the Supporting Information. We note in passing
that often the decay exponents for the computed binding curves deviate from the ideal
values, although the general trend of faster decay when moving from CC to AA dimers is
preserved. Even though this dataset is very similar to that used in
Refs.~\citenum{grisafi_incorporating_2019, grisafi_multi-scale_2021}, we perform an
experiment geared more explicitly towards probing the ability of LR models to capture
the tails of different types of interactions. We split the structures into train and
test sets based on a threshold separation distance \rev{$r_\mathrm{train}$ measured from
the shortest separation $r_0$}, i.e. we train on shorter-range information and assess
whether the model can extrapolate the asymptotic decay. In addition, we include the
dissociated limit of vanishing interaction energy, where the supercell contains only one
monomer. This setup is consistent with \rev{typical scenarios, in which one would like
to train the ML models on small simulation cells and use them on larger structures which
are inaccessible to fully quantum mechanical methods due to the high computational
cost}.

As a first experiment, we fit each dimer class separately, using linear models based on
a multiscale power spectrum\cite{grisafi_multi-scale_2021} that consists of suitable
rotationally invariant products combining both SR descriptors and LODE \rev{using a
single exponent in the range} $p = 1,2, \dots 9$. The linear model details and fitting
procedures are discussed further in \autoref{si-sec:linear_model_dimers} of the
Supporting Information. We analyze the performance of the resulting models in
\autoref{fig:rmse_subset}, for a training threshold $r_\mathrm{train}=4$\,Å. The
qualitative observations for other threshold distances, shown in the Supporting
Information (\autoref{si-fig:dimers_learning_curve}) are similar.
\autoref{fig:rmse_subset}a shows the test-set energy RMSE as a function of the model's
potential exponent $p$ for each class of dimers, using \rev{an optimized radial basis
for $l_\mathrm{max}=1$}. In all cases, SR models show poor performance, \rev{while the
LODE models with appropriate exponents can lead to dramatic improvements by up to an
order of magnitude.} For example, for the CC pairs we find the best model for $p=1$,
while $p=6$ perform best for the AA pairs. The optimal values roughly correspond to the
ideal exponents for each dimer class. The fact that the variation is somewhat smooth can
be understood based on the fact that (\rev{1) the model potential is built as a
superposition of atomic contributions. Over a finite distance interval, any $1/r^p$
potential can be fitted reasonably well with a superposition of multiple
$1/\|\br-\br_i\|^{p'}$ potentials with origins shifted to the atomic positions, even if
the exponents $p\neq p'$ do not agree} (see \autoref{si-sec:fitting_exponents} in the in
the Supporting Information for a more in-depth discussion); (2) binding curves for real
molecules do not exactly match the ideal behavior; (3) interactions with $p>1$ require
more than a single monomial basis. Indeed, considering multiple monomial basis functions
($n_\mathrm{max}=6$, $l_\mathrm{max}=4$) changes the performance curves
(\autoref{fig:rmse_subset}b). There is a large improvement for dimer classes with a
large $p$ (which is consistent with significant contributions from several monomial
coefficients), but also for $p=1$. The dependency of the accuracy on the LODE exponent
is also less sharp, which is consistent with our theoretical analysis, that shows that
higher-$l$ basis functions lead to terms that decay as $1/r^{p+l}$. \rev{It should be
noted that for the best exponents, the results are best for the AA and PA subsets, which
is in part due to the larger size of the training set (around sixfold for AA compared to
CC). In addition, the training is done on the total energy and forces, which also
include SR terms and prove difficult to fit with this data set that is entirely focused
on long-range contributions.} This comparison underscores a key observation in this
work: LR interactions in realistic systems cannot be fully captured by the idealized
asymptotics. Still, LODE features with an exponent adapted to the dimer class usually
\emph{do} perform better, which testifies to the added value of using a
physically-interpretable framework.

\begin{figure}
    \includegraphics[width=0.49\textwidth]{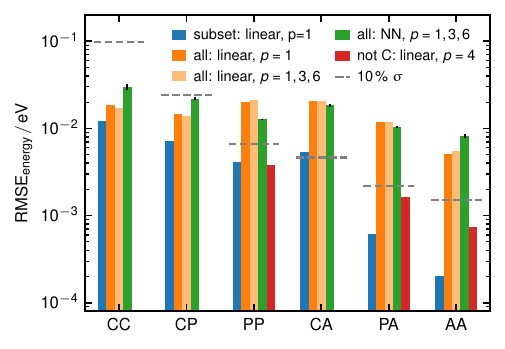}
    \caption{Energy RMSEs for the different subsets of the dimers. Blue bars show the
        RMSEs of models using a single $p=1$ LODE exponent similar to
        \autoref{fig:rmse_subset}a, but \rev{fitted only on energies. Orange and light
        orange bars show linear models fit to the whole dataset, either only using $p=1$
        or combining $p=1,3,6$ for the features. Green bars show a fit to the whole
        dataset with a non-linear neural network model, again with features combining
        $p=1,3,6$. Red bars correspond to a linear model restricted to non-charged
        fragments, using $p=4$ (see \autoref{si-sec:linear_model_dimers} of the
        Supporting Information). Horizontal gray lines depict an relative error of
        $10\%\,\sigma$ for each subset.}}
    \label{fig:all_comparison}
\end{figure}

\rev{Given the ability of LODE features to target different types of interactions, it is
interesting to investigate a model trained against the entire dataset}. This is
particularly challenging because of the vastly different energy scale of the
interactions, which is apparent in \rev{\autoref{fig:rmse_subset},} and because it
is more difficult for the model to infer the correct asymptotic behavior of the
interactions based on SR training data. A linear model based only on $p=1$ coefficients
yields respectable performance for all dimer classes \emph{separately}, but when applied
to the full dataset the errors increase by an order of magnitude or more
(\autoref{fig:all_comparison}). \rev{When compared with the intrinsic energy scale,}
results are particularly poor for the faster-decaying interactions that are completely
overshadowed by the much stronger variability in the CC binding curves: in terms of
absolute errors, the error on the LR part for PP, PA, AA dimers is comparable to that on
the CC subset.

The overwhelming dominance of strong interactions cannot be addressed by  using more
flexible models that include multiple exponents, nor by adding a non-linear layer on top
of the LODE features \rev{(\autoref{fig:all_comparison}, light orange and green
bars).} A more effective strategy, instead, is to reduce the variability in energy scale
by restricting the model to the dimer types that do not contain charged residues
(\autoref{fig:all_comparison}, red bars). This reduces the errors on PP, PA and AA
dimers by an order of magnitude. Larger training sets might allow to increase the
accuracy when targeting all types of non-bonded interactions. However, our computational
experiment highlights one of the inherent challenges when extending ML potential to LR
physics. Without including ad hoc structxures, and designing training targets that
single out the desired type of contributions, fitting models on total energy and forces
focuses on the larger, SR or electrostatic terms. Other non-bonded interactions that are
important to drive collective effects, but small in absolute magnitude, risk being
completely neglected even with a model based on physically-inspired terms.

Summarizing our results, we present an extension of the long-distance equivariant
framework to arbitrary potential exponents,  and give a direct physical interpretation
of the LODE coefficients. 
\rev{ We prove a direct link between LODE features and the multipole expansion -- which we use
to propose a physically-motivated radial basis which is adapted to the modeling of the
far-field contributions -- and show that combining LODE coefficients in a way analogous to what is done with atom-centered density correlations probvides a systematic way to express many-body long-range physics.}
We also provide a fast and modular implementation of this
framework using the GTO as well as an optimized radial basis, which also includes
calculations of gradients, making it possible to train on forces. This
physically-interpretable yet generally-applicable class of descriptors allows us to
investigate the challenges inherent in the ML modeling of LR interactions, revealing the
delicate balance between physical content and generality of a model. While it is
beneficial to use descriptors that are designed to reflect the expected asymptotic decay
of the interactions, we observe that in a realistic application doing so is neither
strictly necessary, nor a guarantee of success. 
\rev{LODE-based models can reach 
levels of accuracy of a small fraction of the typical binding energy for different types of asymptotic behavior, but only when separately targeting  chemically homogeneous sets of molecules.} 
When considering a heterogeneous dataset that contains
very different kinds of intermolecular interactions, the large variability in the energy
scale of the different physical terms makes it particularly difficult to achieve good
relative accuracy for the weaker types of interactions. The description of LR
interactions, from the bare interactions themselves to more complex processes such as
nonlocal charge transfers, remains one of the key challenges to the application of ML
techniques to atomistic simulations. The extension of the LODE framework we introduce
here provides a flexible, physically-motivated solution, and a sandbox to investigate
the effects of descriptor engineering and ML architecture, balancing interpretability
and generality of the model.

\section*{Acknowledgements}
    The Authors acknowledge funding from the European Research Council (ERC) under the
    European Union’s Horizon 2020 research and innovation programme (grant agreement No
    101001890-FIAMMA). We would like to thank Guillaume Fraux and all the members of the
    Laboratory of Computational Science and Modeling for their contributions to the
    software infrastructure that enabled this study. We also thank Andrea Grisafi and
    Jigyasa Nigam for insightful discussion.

\section*{Supporting material}
    The electronic supporting material for this publication include a pedagogic
    derivation of the connection between generalized LODE features and the multipole
    expansion, further details on the different models and datasets, and the dimer
    dataset used for the benchmarks. All used datasets as well as the input for the DFT
    calculations are available for download at
    \url{https://doi.org/10.24435/materialscloud:23-99}. Generalized LODE descriptors
    can be computed using the \emph{rascaline} package, available at
    \url{https://github.com/luthaf/rascaline}. Additional source codes for constructing
    the monomial basis in \emph{rascaline} as well as the code to train the linear and
    the neural networks are available on zenodo
    \url{https://doi.org/10.5281/zenodo.8399545}

\providecommand{\noopsort}[1]{}
\providecommand{\latin}[1]{#1}
\makeatletter
\providecommand{\doi}
  {\begingroup\let\do\@makeother\dospecials
  \catcode`\{=1 \catcode`\}=2 \doi@aux}
\providecommand{\doi@aux}[1]{\endgroup\texttt{#1}}
\makeatother
\providecommand*\mcitethebibliography{\thebibliography}
\csname @ifundefined\endcsname{endmcitethebibliography}
  {\let\endmcitethebibliography\endthebibliography}{}

\vfill
\newcounter{sipage}
\setcounter{sipage}{1}
\loop
{%
\clearpage
\includepdf[pages={\thesipage}]{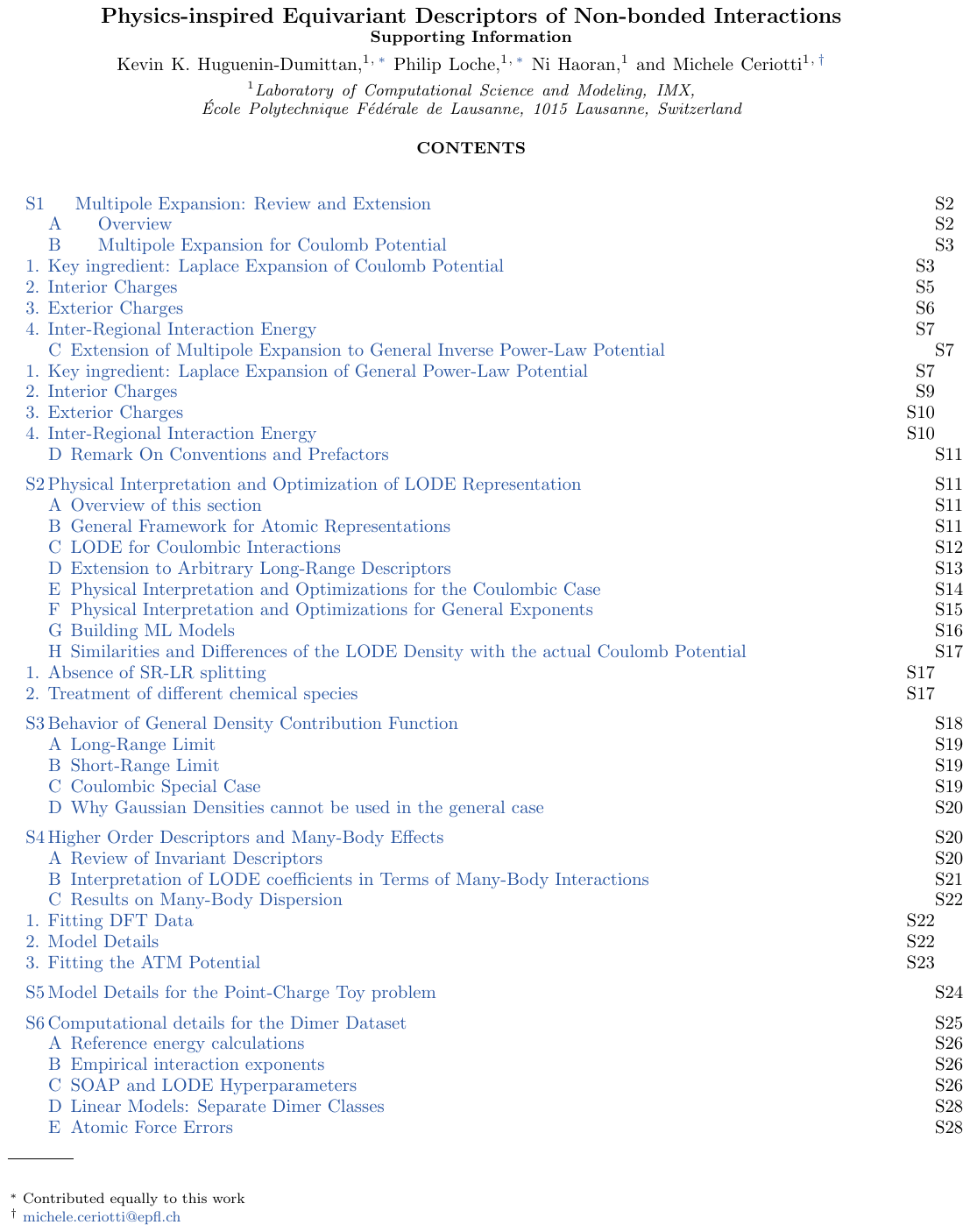}
}
\addtocounter{sipage}{1}
\ifnum \value{sipage}<34
\repeat


\begin{mcitethebibliography}{51}
\providecommand*\natexlab[1]{#1}
\providecommand*\mciteSetBstSublistMode[1]{}
\providecommand*\mciteSetBstMaxWidthForm[2]{}
\providecommand*\mciteBstWouldAddEndPuncttrue
  {\def\EndOfBibitem{\unskip.}}
\providecommand*\mciteBstWouldAddEndPunctfalse
  {\let\EndOfBibitem\relax}
\providecommand*\mciteSetBstMidEndSepPunct[3]{}
\providecommand*\mciteSetBstSublistLabelBeginEnd[3]{}
\providecommand*\EndOfBibitem{}
\mciteSetBstSublistMode{f}
\mciteSetBstMaxWidthForm{subitem}{(\alph{mcitesubitemcount})}
\mciteSetBstSublistLabelBeginEnd
  {\mcitemaxwidthsubitemform\space}
  {\relax}
  {\relax}

\bibitem[Behler and Parrinello(2007)Behler, and
  Parrinello]{behler_generalized_2007}
Behler,~J.; Parrinello,~M. Generalized {{Neural-Network Representation}} of
  {{High-Dimensional Potential-Energy Surfaces}}. \emph{Phys. Rev. Lett.}
  \textbf{2007}, \emph{98}, 146401\relax
\mciteBstWouldAddEndPuncttrue
\mciteSetBstMidEndSepPunct{\mcitedefaultmidpunct}
{\mcitedefaultendpunct}{\mcitedefaultseppunct}\relax
\EndOfBibitem
\bibitem[Bart{\'o}k \latin{et~al.}(2010)Bart{\'o}k, Payne, Kondor, and
  Cs{\'a}nyi]{bartok_gaussian_2010}
Bart{\'o}k,~A.~P.; Payne,~M.~C.; Kondor,~R.; Cs{\'a}nyi,~G. Gaussian
  {{Approximation Potentials}}: {{The Accuracy}} of {{Quantum Mechanics}},
  without the {{Electrons}}. \emph{Phys. Rev. Lett.} \textbf{2010}, \emph{104},
  136403\relax
\mciteBstWouldAddEndPuncttrue
\mciteSetBstMidEndSepPunct{\mcitedefaultmidpunct}
{\mcitedefaultendpunct}{\mcitedefaultseppunct}\relax
\EndOfBibitem
\bibitem[Bart{\'o}k \latin{et~al.}(2013)Bart{\'o}k, Kondor, and
  Cs{\'a}nyi]{bartok_representing_2013}
Bart{\'o}k,~A.~P.; Kondor,~R.; Cs{\'a}nyi,~G. On Representing Chemical
  Environments. \emph{Phys. Rev. B} \textbf{2013}, \emph{87}, 184115\relax
\mciteBstWouldAddEndPuncttrue
\mciteSetBstMidEndSepPunct{\mcitedefaultmidpunct}
{\mcitedefaultendpunct}{\mcitedefaultseppunct}\relax
\EndOfBibitem
\bibitem[Musil \latin{et~al.}(2021)Musil, Grisafi, Bart{\'o}k, Ortner,
  Cs{\'a}nyi, and Ceriotti]{musil_physics-inspired_2021}
Musil,~F.; Grisafi,~A.; Bart{\'o}k,~A.~P.; Ortner,~C.; Cs{\'a}nyi,~G.;
  Ceriotti,~M. Physics-{{Inspired Structural Representations}} for
  {{Molecules}} and {{Materials}}. \emph{Chem. Rev.} \textbf{2021}, \emph{121},
  9759--9815\relax
\mciteBstWouldAddEndPuncttrue
\mciteSetBstMidEndSepPunct{\mcitedefaultmidpunct}
{\mcitedefaultendpunct}{\mcitedefaultseppunct}\relax
\EndOfBibitem
\bibitem[Rupp \latin{et~al.}(2012)Rupp, Tkatchenko, M{\"u}ller, and
  {\noopsort{lilienfeld}}{von Lilienfeld}]{rupp+12prl}
Rupp,~M.; Tkatchenko,~A.; M{\"u}ller,~K.-R.; {\noopsort{lilienfeld}}{von
  Lilienfeld},~O.~A. Fast and {{Accurate Modeling}} of {{Molecular Atomization
  Energies}} with {{Machine Learning}}. \emph{Phys. Rev. Lett.} \textbf{2012},
  \emph{108}, 058301\relax
\mciteBstWouldAddEndPuncttrue
\mciteSetBstMidEndSepPunct{\mcitedefaultmidpunct}
{\mcitedefaultendpunct}{\mcitedefaultseppunct}\relax
\EndOfBibitem
\bibitem[Glielmo \latin{et~al.}(2017)Glielmo, Sollich, and De~Vita]{glie+17prb}
Glielmo,~A.; Sollich,~P.; De~Vita,~A. Accurate Interatomic Force Fields via
  Machine Learning with Covariant Kernels. \emph{Phys. Rev. B} \textbf{2017},
  \emph{95}, 214302\relax
\mciteBstWouldAddEndPuncttrue
\mciteSetBstMidEndSepPunct{\mcitedefaultmidpunct}
{\mcitedefaultendpunct}{\mcitedefaultseppunct}\relax
\EndOfBibitem
\bibitem[Grisafi \latin{et~al.}(2018)Grisafi, Wilkins, Cs{\'a}nyi, and
  Ceriotti]{grisafi_symmetry-adapted_2018}
Grisafi,~A.; Wilkins,~D.~M.; Cs{\'a}nyi,~G.; Ceriotti,~M. Symmetry-{{Adapted
  Machine Learning}} for {{Tensorial Properties}} of {{Atomistic Systems}}.
  \emph{Phys. Rev. Lett.} \textbf{2018}, \emph{120}, 036002\relax
\mciteBstWouldAddEndPuncttrue
\mciteSetBstMidEndSepPunct{\mcitedefaultmidpunct}
{\mcitedefaultendpunct}{\mcitedefaultseppunct}\relax
\EndOfBibitem
\bibitem[Wilkins \latin{et~al.}(2019)Wilkins, Grisafi, Yang, Lao, DiStasio, and
  Ceriotti]{wilkins_accurate_2019}
Wilkins,~D.~M.; Grisafi,~A.; Yang,~Y.; Lao,~K.~U.; DiStasio,~R.~A.;
  Ceriotti,~M. Accurate Molecular Polarizabilities with Coupled Cluster Theory
  and Machine Learning. \emph{PNAS} \textbf{2019}, \emph{116}, 3401--3406\relax
\mciteBstWouldAddEndPuncttrue
\mciteSetBstMidEndSepPunct{\mcitedefaultmidpunct}
{\mcitedefaultendpunct}{\mcitedefaultseppunct}\relax
\EndOfBibitem
\bibitem[Brockherde \latin{et~al.}(2017)Brockherde, Vogt, Li, Tuckerman, Burke,
  and M{\"u}ller]{broc+17nc}
Brockherde,~F.; Vogt,~L.; Li,~L.; Tuckerman,~M.~E.; Burke,~K.;
  M{\"u}ller,~K.~R. Bypassing the {{Kohn-Sham}} Equations with Machine
  Learning. \emph{Nat. Commun.} \textbf{2017}, \emph{8}, 872\relax
\mciteBstWouldAddEndPuncttrue
\mciteSetBstMidEndSepPunct{\mcitedefaultmidpunct}
{\mcitedefaultendpunct}{\mcitedefaultseppunct}\relax
\EndOfBibitem
\bibitem[Alred \latin{et~al.}(2018)Alred, Bets, Xie, and Yakobson]{alre+18cst}
Alred,~J.~M.; Bets,~K.~V.; Xie,~Y.; Yakobson,~B.~I. Machine Learning Electron
  Density in Sulfur Crosslinked Carbon Nanotubes. \emph{Composites Science and
  Technology} \textbf{2018}, \emph{166}, 3--9\relax
\mciteBstWouldAddEndPuncttrue
\mciteSetBstMidEndSepPunct{\mcitedefaultmidpunct}
{\mcitedefaultendpunct}{\mcitedefaultseppunct}\relax
\EndOfBibitem
\bibitem[Fabrizio \latin{et~al.}(2019)Fabrizio, Grisafi, Meyer, Ceriotti, and
  Corminboeuf]{fabrizio_electron_2019}
Fabrizio,~A.; Grisafi,~A.; Meyer,~B.; Ceriotti,~M.; Corminboeuf,~C. Electron
  Density Learning of Non-Covalent Systems. \emph{Chemical Science}
  \textbf{2019}, \emph{10}, 9424--9432\relax
\mciteBstWouldAddEndPuncttrue
\mciteSetBstMidEndSepPunct{\mcitedefaultmidpunct}
{\mcitedefaultendpunct}{\mcitedefaultseppunct}\relax
\EndOfBibitem
\bibitem[Lewis \latin{et~al.}(2021)Lewis, Grisafi, Ceriotti, and
  Rossi]{lewis_learning_2021}
Lewis,~A.~M.; Grisafi,~A.; Ceriotti,~M.; Rossi,~M. Learning {{Electron
  Densities}} in the {{Condensed Phase}}. \emph{J. Chem. Theory Comput.}
  \textbf{2021}, \emph{17}, 7203--7214\relax
\mciteBstWouldAddEndPuncttrue
\mciteSetBstMidEndSepPunct{\mcitedefaultmidpunct}
{\mcitedefaultendpunct}{\mcitedefaultseppunct}\relax
\EndOfBibitem
\bibitem[Sch{\"u}tt \latin{et~al.}(2019)Sch{\"u}tt, Gastegger, Tkatchenko,
  M{\"u}ller, and Maurer]{schu+19nc}
Sch{\"u}tt,~K.~T.; Gastegger,~M.; Tkatchenko,~A.; M{\"u}ller,~K.-R.;
  Maurer,~R.~J. Unifying Machine Learning and Quantum Chemistry with a Deep
  Neural Network for Molecular Wavefunctions. \emph{Nat Commun} \textbf{2019},
  \emph{10}, 5024\relax
\mciteBstWouldAddEndPuncttrue
\mciteSetBstMidEndSepPunct{\mcitedefaultmidpunct}
{\mcitedefaultendpunct}{\mcitedefaultseppunct}\relax
\EndOfBibitem
\bibitem[Nigam \latin{et~al.}(2022)Nigam, Willatt, and
  Ceriotti]{nigam_equivariant_2022}
Nigam,~J.; Willatt,~M.~J.; Ceriotti,~M. Equivariant Representations for
  Molecular {{Hamiltonians}} and {{N-center}} Atomic-Scale Properties. \emph{J.
  Chem. Phys.} \textbf{2022}, \emph{156}, 014115\relax
\mciteBstWouldAddEndPuncttrue
\mciteSetBstMidEndSepPunct{\mcitedefaultmidpunct}
{\mcitedefaultendpunct}{\mcitedefaultseppunct}\relax
\EndOfBibitem
\bibitem[Carleo and Troyer(2017)Carleo, and Troyer]{carl-troy17science}
Carleo,~G.; Troyer,~M. Solving the Quantum Many-Body Problem with Artificial
  Neural Networks. \emph{Science} \textbf{2017}, \emph{355}, 602--606\relax
\mciteBstWouldAddEndPuncttrue
\mciteSetBstMidEndSepPunct{\mcitedefaultmidpunct}
{\mcitedefaultendpunct}{\mcitedefaultseppunct}\relax
\EndOfBibitem
\bibitem[Hermann \latin{et~al.}(2020)Hermann, Sch{\"a}tzle, and
  No{\'e}]{herm+20nc}
Hermann,~J.; Sch{\"a}tzle,~Z.; No{\'e},~F. Deep-Neural-Network Solution of the
  Electronic {{Schr\"odinger}} Equation. \emph{Nat. Chem.} \textbf{2020},
  \emph{12}, 891--897\relax
\mciteBstWouldAddEndPuncttrue
\mciteSetBstMidEndSepPunct{\mcitedefaultmidpunct}
{\mcitedefaultendpunct}{\mcitedefaultseppunct}\relax
\EndOfBibitem
\bibitem[Kohn(1996)]{kohn_density_1996}
Kohn,~W. Density Functional and Density Matrix Method Scaling Linearly with the
  Number of Atoms. \emph{Phys. Rev. Lett.} \textbf{1996}, \emph{76},
  3168--3171\relax
\mciteBstWouldAddEndPuncttrue
\mciteSetBstMidEndSepPunct{\mcitedefaultmidpunct}
{\mcitedefaultendpunct}{\mcitedefaultseppunct}\relax
\EndOfBibitem
\bibitem[Prodan and Kohn(2005)Prodan, and Kohn]{prodan_nearsightedness_2005}
Prodan,~E.; Kohn,~W. Nearsightedness of Electronic Matter. \emph{Proceedings of
  the National Academy of Sciences} \textbf{2005}, \emph{102},
  11635--11638\relax
\mciteBstWouldAddEndPuncttrue
\mciteSetBstMidEndSepPunct{\mcitedefaultmidpunct}
{\mcitedefaultendpunct}{\mcitedefaultseppunct}\relax
\EndOfBibitem
\bibitem[Ambrosetti \latin{et~al.}(2016)Ambrosetti, Ferri, DiStasio, and
  Tkatchenko]{ambr+16science}
Ambrosetti,~A.; Ferri,~N.; DiStasio,~R.~A.; Tkatchenko,~A. Wavelike Charge
  Density Fluctuations and van Der {{Waals}} Interactions at the Nanoscale.
  \emph{Science} \textbf{2016}, \emph{351}, 1171--1176\relax
\mciteBstWouldAddEndPuncttrue
\mciteSetBstMidEndSepPunct{\mcitedefaultmidpunct}
{\mcitedefaultendpunct}{\mcitedefaultseppunct}\relax
\EndOfBibitem
\bibitem[Jackson(1998)]{jackson_classical_1998}
Jackson,~J.~D. \emph{Classical {{Electrodynamics Third Edition}}}, 3rd ed.;
  {Wiley}: {New York}, 1998\relax
\mciteBstWouldAddEndPuncttrue
\mciteSetBstMidEndSepPunct{\mcitedefaultmidpunct}
{\mcitedefaultendpunct}{\mcitedefaultseppunct}\relax
\EndOfBibitem
\bibitem[Israelachvili(2011)]{israelachvili_intermolecular_2011}
Israelachvili,~J.~N. \emph{Intermolecular and {{Surface Forces}} ({{Third
  Edition}})}, third edition ed.; {Academic Press}: {San Diego}, 2011\relax
\mciteBstWouldAddEndPuncttrue
\mciteSetBstMidEndSepPunct{\mcitedefaultmidpunct}
{\mcitedefaultendpunct}{\mcitedefaultseppunct}\relax
\EndOfBibitem
\bibitem[Kandu{\v c} \latin{et~al.}(2014)Kandu{\v c}, Schneck, and
  Netz]{kanduc_attraction_2014}
Kandu{\v c},~M.; Schneck,~E.; Netz,~R.~R. Attraction between Hydrated
  Hydrophilic Surfaces. \emph{Chemical Physics Letters} \textbf{2014},
  \emph{610--611}, 375--380\relax
\mciteBstWouldAddEndPuncttrue
\mciteSetBstMidEndSepPunct{\mcitedefaultmidpunct}
{\mcitedefaultendpunct}{\mcitedefaultseppunct}\relax
\EndOfBibitem
\bibitem[han(2013)]{hansen_theory_2013}
Hansen,~J.-P., McDonald,~I.~R., Eds. \emph{Theory of {{Simple Liquids}}
  ({{Fourth Edition}})}; {Academic Press}: {Oxford}, 2013; p~i\relax
\mciteBstWouldAddEndPuncttrue
\mciteSetBstMidEndSepPunct{\mcitedefaultmidpunct}
{\mcitedefaultendpunct}{\mcitedefaultseppunct}\relax
\EndOfBibitem
\bibitem[Deng \latin{et~al.}(2019)Deng, Chen, Li, and
  Ong]{deng_electrostatic_2019}
Deng,~Z.; Chen,~C.; Li,~X.-G.; Ong,~S.~P. An Electrostatic Spectral Neighbor
  Analysis Potential for Lithium Nitride. \emph{npj Computational Materials}
  \textbf{2019}, \emph{5}, 1--8\relax
\mciteBstWouldAddEndPuncttrue
\mciteSetBstMidEndSepPunct{\mcitedefaultmidpunct}
{\mcitedefaultendpunct}{\mcitedefaultseppunct}\relax
\EndOfBibitem
\bibitem[Deringer \latin{et~al.}(2020)Deringer, Caro, and
  Cs{\'a}nyi]{deringer_general-purpose_2020}
Deringer,~V.~L.; Caro,~M.~A.; Cs{\'a}nyi,~G. A General-Purpose Machine-Learning
  Force Field for Bulk and Nanostructured Phosphorus. \emph{Nat Commun}
  \textbf{2020}, \emph{11}, 5461\relax
\mciteBstWouldAddEndPuncttrue
\mciteSetBstMidEndSepPunct{\mcitedefaultmidpunct}
{\mcitedefaultendpunct}{\mcitedefaultseppunct}\relax
\EndOfBibitem
\bibitem[Niblett \latin{et~al.}(2021)Niblett, Galib, and
  Limmer]{niblett_learning_2021}
Niblett,~S.~P.; Galib,~M.; Limmer,~D.~T. Learning Intermolecular Forces at
  Liquid\textendash Vapor Interfaces. \emph{J. Chem. Phys.} \textbf{2021},
  \emph{155}, 164101\relax
\mciteBstWouldAddEndPuncttrue
\mciteSetBstMidEndSepPunct{\mcitedefaultmidpunct}
{\mcitedefaultendpunct}{\mcitedefaultseppunct}\relax
\EndOfBibitem
\bibitem[Peng \latin{et~al.}(2023)Peng, Lin, Ying, and
  {Zepeda-N{\'u}{\~n}ez}]{peng_efficient_2023}
Peng,~Y.; Lin,~L.; Ying,~L.; {Zepeda-N{\'u}{\~n}ez},~L. Efficient Long-Range
  Convolutions for Point Clouds. \emph{Journal of Computational Physics}
  \textbf{2023}, \emph{473}, 111692\relax
\mciteBstWouldAddEndPuncttrue
\mciteSetBstMidEndSepPunct{\mcitedefaultmidpunct}
{\mcitedefaultendpunct}{\mcitedefaultseppunct}\relax
\EndOfBibitem
\bibitem[Zhang \latin{et~al.}(2022)Zhang, Wang, Muniz, Panagiotopoulos, Car,
  and E]{zhang_deep_2022}
Zhang,~L.; Wang,~H.; Muniz,~M.~C.; Panagiotopoulos,~A.~Z.; Car,~R.; E,~W. A
  Deep Potential Model with Long-Range Electrostatic Interactions. \emph{J.
  Chem. Phys.} \textbf{2022}, \emph{156}, 124107\relax
\mciteBstWouldAddEndPuncttrue
\mciteSetBstMidEndSepPunct{\mcitedefaultmidpunct}
{\mcitedefaultendpunct}{\mcitedefaultseppunct}\relax
\EndOfBibitem
\bibitem[Faraji \latin{et~al.}(2017)Faraji, Ghasemi, Rostami, Rasoulkhani,
  Schaefer, Goedecker, and Amsler]{faraji_high_2017}
Faraji,~S.; Ghasemi,~S.~A.; Rostami,~S.; Rasoulkhani,~R.; Schaefer,~B.;
  Goedecker,~S.; Amsler,~M. High Accuracy and Transferability of a Neural
  Network Potential through Charge Equilibration for Calcium Fluoride.
  \emph{Phys. Rev. B} \textbf{2017}, \emph{95}, 104105\relax
\mciteBstWouldAddEndPuncttrue
\mciteSetBstMidEndSepPunct{\mcitedefaultmidpunct}
{\mcitedefaultendpunct}{\mcitedefaultseppunct}\relax
\EndOfBibitem
\bibitem[Ko \latin{et~al.}(2021)Ko, Finkler, Goedecker, and
  Behler]{ko_fourth-generation_2021}
Ko,~T.~W.; Finkler,~J.~A.; Goedecker,~S.; Behler,~J. A Fourth-Generation
  High-Dimensional Neural Network Potential with Accurate Electrostatics
  Including Non-Local Charge Transfer. \emph{Nat Commun} \textbf{2021},
  \emph{12}, 398\relax
\mciteBstWouldAddEndPuncttrue
\mciteSetBstMidEndSepPunct{\mcitedefaultmidpunct}
{\mcitedefaultendpunct}{\mcitedefaultseppunct}\relax
\EndOfBibitem
\bibitem[Ko \latin{et~al.}(2021)Ko, Finkler, Goedecker, and
  Behler]{ko_general-purpose_2021}
Ko,~T.~W.; Finkler,~J.~A.; Goedecker,~S.; Behler,~J. General-{{Purpose Machine
  Learning Potentials Capturing Nonlocal Charge Transfer}}. \emph{Acc. Chem.
  Res.} \textbf{2021}, \emph{54}, 808--817\relax
\mciteBstWouldAddEndPuncttrue
\mciteSetBstMidEndSepPunct{\mcitedefaultmidpunct}
{\mcitedefaultendpunct}{\mcitedefaultseppunct}\relax
\EndOfBibitem
\bibitem[Gao and Remsing(2022)Gao, and Remsing]{gao-rems22nc}
Gao,~A.; Remsing,~R.~C. Self-Consistent Determination of Long-Range
  Electrostatics in Neural Network Potentials. \emph{Nat Commun} \textbf{2022},
  \emph{13}, 1572\relax
\mciteBstWouldAddEndPuncttrue
\mciteSetBstMidEndSepPunct{\mcitedefaultmidpunct}
{\mcitedefaultendpunct}{\mcitedefaultseppunct}\relax
\EndOfBibitem
\bibitem[Anstine and Isayev(2023)Anstine, and Isayev]{anstine_machine_2023}
Anstine,~D.~M.; Isayev,~O. Machine {{Learning Interatomic Potentials}} and
  {{Long-Range Physics}}. \emph{J. Chem. Phys.} \textbf{2023}, \emph{127},
  2417--2431\relax
\mciteBstWouldAddEndPuncttrue
\mciteSetBstMidEndSepPunct{\mcitedefaultmidpunct}
{\mcitedefaultendpunct}{\mcitedefaultseppunct}\relax
\EndOfBibitem
\bibitem[Grisafi and Ceriotti(2019)Grisafi, and
  Ceriotti]{grisafi_incorporating_2019}
Grisafi,~A.; Ceriotti,~M. Incorporating Long-Range Physics in Atomic-Scale
  Machine Learning. \emph{J. Chem. Phys.} \textbf{2019}, \emph{151},
  204105\relax
\mciteBstWouldAddEndPuncttrue
\mciteSetBstMidEndSepPunct{\mcitedefaultmidpunct}
{\mcitedefaultendpunct}{\mcitedefaultseppunct}\relax
\EndOfBibitem
\bibitem[Grisafi \latin{et~al.}(2021)Grisafi, Nigam, and
  Ceriotti]{grisafi_multi-scale_2021}
Grisafi,~A.; Nigam,~J.; Ceriotti,~M. Multi-Scale Approach for the Prediction of
  Atomic Scale Properties. \emph{Chem. Sci.} \textbf{2021}, \emph{12},
  2078--2090\relax
\mciteBstWouldAddEndPuncttrue
\mciteSetBstMidEndSepPunct{\mcitedefaultmidpunct}
{\mcitedefaultendpunct}{\mcitedefaultseppunct}\relax
\EndOfBibitem
\bibitem[Grisafi and Ceriotti(2019)Grisafi, and Ceriotti]{gris-ceri19jcp}
Grisafi,~A.; Ceriotti,~M. Incorporating Long-Range Physics in Atomic-Scale
  Machine Learning. \emph{J. Chem. Phys.} \textbf{2019}, \emph{151},
  204105\relax
\mciteBstWouldAddEndPuncttrue
\mciteSetBstMidEndSepPunct{\mcitedefaultmidpunct}
{\mcitedefaultendpunct}{\mcitedefaultseppunct}\relax
\EndOfBibitem
\bibitem[Grisafi \latin{et~al.}(2021)Grisafi, Nigam, and Ceriotti]{gris+21cs}
Grisafi,~A.; Nigam,~J.; Ceriotti,~M. Multi-Scale Approach for the Prediction of
  Atomic Scale Properties. \emph{Chem. Sci.} \textbf{2021}, \emph{12},
  2078--2090\relax
\mciteBstWouldAddEndPuncttrue
\mciteSetBstMidEndSepPunct{\mcitedefaultmidpunct}
{\mcitedefaultendpunct}{\mcitedefaultseppunct}\relax
\EndOfBibitem
\bibitem[Nijboer and De~Wette(1957)Nijboer, and
  De~Wette]{nijboer_calculation_1957}
Nijboer,~B. R.~A.; De~Wette,~F.~W. On the Calculation of Lattice Sums.
  \emph{Physica} \textbf{1957}, \emph{23}, 309--321\relax
\mciteBstWouldAddEndPuncttrue
\mciteSetBstMidEndSepPunct{\mcitedefaultmidpunct}
{\mcitedefaultendpunct}{\mcitedefaultseppunct}\relax
\EndOfBibitem
\bibitem[Willatt \latin{et~al.}(2019)Willatt, Musil, and Ceriotti]{will+19jcp}
Willatt,~M.~J.; Musil,~F.; Ceriotti,~M. Atom-Density Representations for
  Machine Learning. \emph{J. Chem. Phys.} \textbf{2019}, \emph{150},
  154110\relax
\mciteBstWouldAddEndPuncttrue
\mciteSetBstMidEndSepPunct{\mcitedefaultmidpunct}
{\mcitedefaultendpunct}{\mcitedefaultseppunct}\relax
\EndOfBibitem
\bibitem[Drautz(2019)]{drautz_atomic_2019}
Drautz,~R. Atomic Cluster Expansion for Accurate and Transferable Interatomic
  Potentials. \emph{Phys. Rev. B} \textbf{2019}, \emph{99}, 014104\relax
\mciteBstWouldAddEndPuncttrue
\mciteSetBstMidEndSepPunct{\mcitedefaultmidpunct}
{\mcitedefaultendpunct}{\mcitedefaultseppunct}\relax
\EndOfBibitem
\bibitem[Williams(1971)]{williams_accelerated_1971}
Williams,~D.~E. Accelerated Convergence of Crystal-Lattice Potential Sums.
  \emph{Acta Crystallographica Section A: Crystal Physics, Diffraction,
  Theoretical and General Crystallography} \textbf{1971}, \emph{27},
  452--455\relax
\mciteBstWouldAddEndPuncttrue
\mciteSetBstMidEndSepPunct{\mcitedefaultmidpunct}
{\mcitedefaultendpunct}{\mcitedefaultseppunct}\relax
\EndOfBibitem
\bibitem[Williams(1989)]{williams_accelerated_1989}
Williams,~D.~E. Accelerated {{Convergence Treatment}} of {{R}}-n {{Lattice
  Sums}}. \emph{Crystallography Reviews} \textbf{1989}, \emph{2}, 3--23\relax
\mciteBstWouldAddEndPuncttrue
\mciteSetBstMidEndSepPunct{\mcitedefaultmidpunct}
{\mcitedefaultendpunct}{\mcitedefaultseppunct}\relax
\EndOfBibitem
\bibitem[Nigam \latin{et~al.}(2020)Nigam, Pozdnyakov, and Ceriotti]{niga+20jcp}
Nigam,~J.; Pozdnyakov,~S.; Ceriotti,~M. Recursive Evaluation and Iterative
  Contraction of {{{\emph{N}}}} -Body Equivariant Features. \emph{J. Chem.
  Phys.} \textbf{2020}, \emph{153}, 121101\relax
\mciteBstWouldAddEndPuncttrue
\mciteSetBstMidEndSepPunct{\mcitedefaultmidpunct}
{\mcitedefaultendpunct}{\mcitedefaultseppunct}\relax
\EndOfBibitem
\bibitem[Darden \latin{et~al.}(1993)Darden, York, and
  Pedersen]{darden_particle_1993}
Darden,~T.; York,~D.; Pedersen,~L. Particle Mesh {{Ewald}}: {{An
  N}}{$\cdot$}log({{N}}) Method for {{Ewald}} Sums in Large Systems. \emph{J.
  Chem. Phys.} \textbf{1993}, \emph{98}, 10089--10092\relax
\mciteBstWouldAddEndPuncttrue
\mciteSetBstMidEndSepPunct{\mcitedefaultmidpunct}
{\mcitedefaultendpunct}{\mcitedefaultseppunct}\relax
\EndOfBibitem
\bibitem[Essmann \latin{et~al.}(1995)Essmann, Perera, Berkowitz, Darden, Lee,
  and Pedersen]{essmann_smooth_1995}
Essmann,~U.; Perera,~L.; Berkowitz,~M.~L.; Darden,~T.; Lee,~H.; Pedersen,~L.~G.
  A Smooth Particle Mesh {{Ewald}} Method. \emph{J. Chem. Phys.} \textbf{1995},
  \emph{103}, 8577--8593\relax
\mciteBstWouldAddEndPuncttrue
\mciteSetBstMidEndSepPunct{\mcitedefaultmidpunct}
{\mcitedefaultendpunct}{\mcitedefaultseppunct}\relax
\EndOfBibitem
\bibitem[Eastwood(1988)]{eastwood_particle-particleparticle-mesh_1988}
Eastwood,~R. W.~H.,~J.~W. \emph{Computer {{Simulation Using Particles}}}; {CRC
  Press}, 1988\relax
\mciteBstWouldAddEndPuncttrue
\mciteSetBstMidEndSepPunct{\mcitedefaultmidpunct}
{\mcitedefaultendpunct}{\mcitedefaultseppunct}\relax
\EndOfBibitem
\bibitem[Frenkel and Smit(2002)Frenkel, and Smit]{frenkel_understanding_2002}
Frenkel,~D.; Smit,~B. \emph{Understanding Molecular Simulation: From Algorithms
  to Applications}, 2nd ed.; Computational Science Series 1; {Academic Press}:
  {San Diego}, 2002\relax
\mciteBstWouldAddEndPuncttrue
\mciteSetBstMidEndSepPunct{\mcitedefaultmidpunct}
{\mcitedefaultendpunct}{\mcitedefaultseppunct}\relax
\EndOfBibitem
\bibitem[Burns \latin{et~al.}(2017)Burns, Faver, Zheng, Marshall, Smith,
  Vanommeslaeghe, MacKerell, Merz, and Sherrill]{burns_biofragment_2017}
Burns,~L.~A.; Faver,~J.~C.; Zheng,~Z.; Marshall,~M.~S.; Smith,~D. G.~A.;
  Vanommeslaeghe,~K.; MacKerell,~A.~D.; Merz,~K.~M.; Sherrill,~C.~D. The
  {{BioFragment Database}} ({{BFDb}}): {{An}} Open-Data Platform for
  Computational Chemistry Analysis of Noncovalent Interactions. \emph{J. Chem.
  Phys.} \textbf{2017}, \emph{147}, 161727\relax
\mciteBstWouldAddEndPuncttrue
\mciteSetBstMidEndSepPunct{\mcitedefaultmidpunct}
{\mcitedefaultendpunct}{\mcitedefaultseppunct}\relax
\EndOfBibitem
\bibitem[Heyd \latin{et~al.}(2003)Heyd, Scuseria, and
  Ernzerhof]{heyd_hybrid_2003}
Heyd,~J.; Scuseria,~G.~E.; Ernzerhof,~M. Hybrid Functionals Based on a Screened
  {{Coulomb}} Potential. \emph{J. Chem. Phys.} \textbf{2003}, \emph{118},
  8207--8215\relax
\mciteBstWouldAddEndPuncttrue
\mciteSetBstMidEndSepPunct{\mcitedefaultmidpunct}
{\mcitedefaultendpunct}{\mcitedefaultseppunct}\relax
\EndOfBibitem
\bibitem[Hermann and Tkatchenko(2020)Hermann, and
  Tkatchenko]{hermann_density_2020}
Hermann,~J.; Tkatchenko,~A. Density {{Functional Model}} for van Der {{Waals
  Interactions}}: {{Unifying Many-Body Atomic Approaches}} with {{Nonlocal
  Functionals}}. \emph{Phys. Rev. Lett.} \textbf{2020}, \emph{124},
  146401\relax
\mciteBstWouldAddEndPuncttrue
\mciteSetBstMidEndSepPunct{\mcitedefaultmidpunct}
{\mcitedefaultendpunct}{\mcitedefaultseppunct}\relax
\EndOfBibitem
\end{mcitethebibliography}
\end{document}